\documentclass[aps,pra,reprint, amsmath, amssymb,superscriptaddress]{revtex4-1}

\usepackage{printlen}
\usepackage{bm}
\usepackage[retainorgcmds]{IEEEtrantools}
\usepackage{graphicx}
\usepackage{float}
\usepackage{mathrsfs}
\usepackage{amsmath}
\usepackage{amssymb}
\usepackage{color}
\usepackage{nicefrac}
\usepackage{epstopdf} 
\usepackage{blindtext}

\newcommand{\ud}{\,\mathrm{d}}

\DeclareMathOperator{\tr}{tr}
\DeclareMathOperator{\im}{Im}
\DeclareMathOperator{\re}{Re}

\begin{document}

\title{Spin-phase-space-entropy production}
\date{\today}
\author{Jader P. Santos}
\affiliation{Universidade Federal do ABC,  09210-580, Santo Andr\'e, Brazil}
\affiliation{Instituto de F\'isica da Universidade de S\~ao Paulo,  05314-970 S\~ao Paulo, Brazil}
\author{Lucas C. C\'eleri}
\affiliation{Instituto de F\'isica, Universidade Federal de Goi\'as, Caixa Postal 131, 74001-970, Goi\^ania, Brazil}
\author{Frederico Brito}
\affiliation{Instituto de F\'isica de S\~ao Carlos, Universidade de S\~ao Paulo, C.P. 369, 13560-970, S\~ao Carlos, SP, Brazil}
\author{Gabriel T. Landi}
\affiliation{Instituto de F\'isica da Universidade de S\~ao Paulo,  05314-970 S\~ao Paulo, Brazil}
%\affiliation{Instituto de F\'isica da Universidade de S\~ao Paulo,  05314-970, S\~ao Paulo, Brazil}
\author{Mauro Paternostro}
\affiliation{Centre for Theoretical Atomic, Molecular and Optical Physics,
School of Mathematics and Physics, Queen's University Belfast, Belfast BT7 1NN, United Kingdom}
%\affiliation{Laboratoire Kastler Brossel, ENS-PSL Research University, 24 rue Lhomond, F-75005 Paris, France}

\begin{abstract}

Quantifying the degree of irreversibility of an  open system dynamics represents a problem of both fundamental and applied relevance. 
Even though a well-known framework exists for thermal baths, the results give diverging results in the limit of zero temperature and are also not readily extended to nonequilibrium reservoirs, such as dephasing baths.
Aimed at filling this gap, in this paper we introduce a phase-space-entropy production framework for quantifying the irreversibility of spin systems undergoing Lindblad dynamics.
The theory is based on the spin Husimi-Q function and its corresponding phase-space entropy, known as Wehrl entropy. 
Unlike the von Neumann entropy production rate, we show that our framework remains valid at any temperature and is also readily extended to arbitrary nonequilibrium baths. 
As an application, we discuss the irreversibility associated with the interaction of a two-level system with a single-photon pulse, a problem which cannot be treated using the conventional approach.

\end{abstract}
\maketitle{}

%%%%%%%%%%%%%%%%%%%%%%%%%%%%%%%%%%%%%%
%
%
%		INTRODUCTION
%
%
%%%%%%%%%%%%%%%%%%%%%%%%%%%%%%%%%%%%%%
\section{\label{sec:int}Introduction}

Irreversible processes undergone by an open system are associated with a production of entropy that is fundamentally different from any possible entropy flows between the system and its environment. 
To separate the two contributions, we usually write the rate of change of the entropy $S$ of a system as 
\begin{equation}\label{dSdt}
\frac{\ud S}{\ud t} = \Pi - \Phi,
\end{equation}
where $\Phi$ is the entropy flux rate from the system to the environment and $\Pi$ is the entropy production rate. 
According to the second law of thermodynamics, we should have $\Pi \geq 0$ and $\Pi = 0$ if and only if the system is in equilibrium. 
Hence, the entropy production rate may be used as a natural quantifier of the degree of irreversibility of a process.
For this reason, a thorough understanding of the entropy production rate is both fundamentally relevant and technologically desirable. 
On the one hand, such understanding would provide the much needed foundation for the emergence of time-symmetry breaking entailed by irreversibility and epitomized, for instance, by seminal results such as Onsager's theory of irreversible currents \cite{Onsager1931,*Onsager1931a,Machlup1953,DeGroot1961,Tisza1957}. 
On the other hand, a characterization of irreversible entropy could help us to design thermodynamically efficient quantum technologies~\cite{Escher2011,Verstraete2008a}.

The description of entropy production in open quantum systems is still an open question, despite substantial progress \cite{Spohn1978,Breuer2003,Deffner2011,Santos2017b,Brunelli2016a,Brunelli2016,Frank2013,Muller-Lennert2013,Audenaert2013,Abe2003,Esposito2010a,Pucci2013,Li2016c,Solano-Carrillo2015,0295-5075-94-3-30001,Oliveira2016a}.
Here we shall be interested in systems described by a master equation of the form 
\begin{equation}\label{M}
\frac{\ud \rho}{\ud t} = - i [H,\rho] + D(\rho),
\end{equation}
where $\rho$ is the system's density matrix,  $H$ is the Hamiltonian, and $D(\rho)$ is the dissipator describing the effects of the bath. 
In Refs. \cite{ALICKI1976249,Alicki1979}, Alicki suggested a relation for the entropy production in terms of the dynamical semigroup $\{\Lambda_t|t\geqslant 0\}$ generated by Eq.~(\ref{M}) and its corresponding invariant state $\Lambda_t\rho^*=\rho^*$.
The relation is given by
\begin{equation} \label{Pi_vN} 
\Pi_\text{vN} = - \frac{\ud }{\ud t} S_\text{vN}(\Lambda_t\rho(0)||\rho^*),
\end{equation}
where $S_\text{vN}(\rho||\sigma) = \tr(\rho \ln \rho - \rho \ln \sigma)$ is the von Neumann relative entropy.
Clearly, with this definition, $\Pi_\text{vN} \geq 0$, with the equality holding only for $\rho = \rho^*$.

In the case of a thermal bath, the invariant state becomes the  Gibbs state, $\rho^* = \rho_\text{eq} = e^{-\beta H}/Z$.
Using Eqs.~(\ref{dSdt}) and (\ref{Pi_vN}), one may then show that in this case, the entropy flux rate  $\Phi_{\text{vN}}$ becomes
\begin{equation}\label{Phi_vN}
\Phi_{\text{vN}} = \frac{\Phi_E}{T} = - \frac{1}{T}\tr\bigg\{H D(\rho)\bigg\},
\end{equation}
which is the familiar Clausius relation between entropy and heat, therefore providing a more physical basis for Eq.~(\ref{Pi_vN}).

However, Eqs.~(\ref{Pi_vN}) and (\ref{Phi_vN}) both diverge in the limit $T\to0$, even though $\ud S/\ud t$ is well behaved.  
Equation~(\ref{Pi_vN}), in particular, diverges whenever the support of $\rho^*$ is not contained in the support of $\rho$ \cite{Abe2003,Audenaert2013}. 
Such divergence has been the subject of substantial investigation \cite{Frank2013,Muller-Lennert2013,Audenaert2013,Abe2003,Esposito2010a,Pucci2013,Lendi1988}, but whether or not it is a physical consequence of the third law of thermodynamics, or merely a mathematical limitation, remains an open question. 
This sets an immediate practical limitation since it renders this  approach inapplicable to any process whose invariant state is pure, therefore excluding several situations typically encountered in the laboratory. 
For instance, it excludes the remarkably simple problem of  spontaneous emission.

%We can also use the definition Eq.~(\ref{Pi_vN}) for evaluate the entropy production rate in the case of a dephasing channel, but we  shall face problems when considering pure states. 

Reference \cite{Santos2017b} has introduced the idea of using phase-space-entropic measures as an alternative to describe irreversibility in open quantum systems. 
As was shown, not only does this fix the above mentioned divergences for pure states, but it also allows for a transparent way of extending the framework to nonequilibrium reservoirs. 
Moreover, it has the advantage of identifying quasiprobability currents in phase space that represent the microscopic manifestations of irreversibility. 
In Ref.~\cite{Santos2017b}, the focus was on Gaussian bosonic states, for which the Wigner function was shown to be an ideal choice, as it is also related to the R\'enyi-2 entropy. 
However, the question of how this formalism could be extended to other systems  was not explored. 

The goal of this work is to derive a theory of entropy production that is applicable to spin systems subject to general reservoirs.  
To achieve this goal, we shall follow a similar approach as in \cite{Santos2017b} and use phase-space techniques based on spin coherent states and the spin Husimi-Q function \cite{1940264}.  
The Husimi function is a quasiprobability distribution commonly used to study the correspondence between quantum and classical dynamics \cite{PhysRevLett.55.645}. 
Among its properties, it is always positive definite. 
This fact was used by Wehrl \cite{Wehrl1978,Wehrl1979,lieb1978,Araki1970} to define a phase-space version of the Shannon entropy. The Wehrl entropy is not a measure of the purity of the wave function as is the von Neumann entropy, but is directly related to the uncertainty area of the Husimi function in phase space \cite{General1995,Buek1995,PhysRevA.54.729}.
For any state, the Wehrl entropy provides an upper bound to the von Neumann entropy, which is saturated only for the case of a coherent state \cite{lieb1978,lieb2014}.
%The Wehrl entropy has also been used before in order to study the entropy production at the quantum level for a closed system \cite{LittleandBigBang}. Here we construct a theory for entropy production for a quantum open spin system, using Husimi function and the Wehrl entropy.

The paper is organized as follows. 
In Sec.~\ref{sec:spin_ps_dynamics}, we present the framework for describing spin systems in  phase space.
We do so using two equivalent approaches, one based on spin coherent states  and the other based on the Schwinger mapping to bosonic systems. 
We thus obtain two definitions for the Husimi function and for the corresponding Wehrl entropy.
In Sec.~\ref{sec:dephasing_channel}, we study the Wehrl entropy production for the dephasing channel and also discuss, as an application, the dynamics of a  spin $1/2$ in a rotating magnetic field.
In Sec.~\ref{sec:amplitude_damping}, we apply our formalism  to the finite-temperature amplitude damping channel and give general expressions for the Wehrl entropy production rate and entropy flux rate, which are valid for any temperature and spin number.  We also show the relation between the Wehrl entropy flux rate and the energy flux rate. 
Explicit results for the spin-$1/2$ case are given as well.
In Sec.~\ref{sec:applications}, we apply these results to the problems of spontaneous emission,  thermal quenches, and a spin $1/2$ in an oscillating magnetic field.
Finally, in Sec.~\ref{sec:fred}, we  study the entropy production of a two-level system interacting with a single-photon pulse.
The conclusions are summarized in Sec.~\ref{sec:conclusions}.
%
%		Spin phase space dynamics
%
%
\section{\label{sec:spin_ps_dynamics} Spin phase space dynamics}

\subsection{Spin coherent state representation}

In this paper, we shall focus on a single spin $J$ with spin operators $J_x$, $J_y$, and $J_z$. 
Instead of working with the density matrix, we approach this problem from a phase-space perspective.
The natural phase-space representation for spin systems is through spin coherent states, which are defined as~\cite{Radcliffe1971}
\begin{equation}\label{spin_coherent_states}
|\Omega\rangle= e^{- i \phi J_z} e^{-i \theta J_y} e^{-i \psi J_z} |J,J\rangle,
\end{equation}
where $|J,J\rangle$ is the angular momentum state with largest quantum number of $J_z$, and $(\theta,\phi,\psi)$ are Euler angles.
The angle $\psi$ is not actually necessary and is placed here only for the sake of completeness.

We may define, as a phase-space distribution for this spin system, the Husimi-Q function,
\begin{equation}\label{spin_Q}
\mathcal{Q}(\Omega) = \langle \Omega | \rho | \Omega \rangle.
\end{equation}
In phase space, the dynamics of Eq.~(\ref{M}) can be recast into the Fokker-Planck equation for $\mathcal{Q}$,
\begin{equation}\label{FP}
\partial_t \mathcal{Q} = U(\mathcal{Q}) +  \mathcal{D} (\mathcal{Q}),
\end{equation}
where $U$ accounts for the unitary part of the evolution and $\mathcal{D}$ for the dissipator.
The phase-space differential operators $U(\mathcal{Q})$ and $\mathcal{D}(\mathcal{Q})$ may then be obtained from standard operator correspondence tables. 
The most interesting correspondences are those concerning commutators of the spin operators $J_i$, which translate into the usual orbital angular momentum operators:
\begin{IEEEeqnarray}{rCl}
\label{Jp_orbital}
[J_+, \rho] &\to& \mathcal{J}_+(\mathcal{Q})   =  e^{i\phi}(\partial_\theta+i\cot\theta\partial_\phi) \mathcal{Q},
\\[0.2cm]
\label{Jm_orbital}
[J_-,\rho] &\to& \mathcal{J}_{-}(\mathcal{Q}) =  -e^{-i\phi}(\partial_\theta-i\cot\theta\partial_\phi) \mathcal{Q},
\\[0.2cm]
\label{Jz_orbital}
[J_z, \rho] &\to& \mathcal{J}_{z}(\mathcal{Q}) = -i \frac{\partial}{\partial \phi} \mathcal{Q}.
\IEEEeqnarraynumspace
\end{IEEEeqnarray}

\subsection{Takahashi-Shibata-Schwinger representation}

Working with spin coherent states can eventually be cumbersome as they do not have the simplicity of standard coherent states.
Here we shall also use a different approach put forth by Takahashi and Shibata \cite{Takahashi1975}, which consists of first using the Schwinger operators to map the spin operators into two bosonic modes and then defining standard phase-space measures using bosonic coherent states. 
We shall thus refer to this as the Takahashi-Shibata-Schwinger (TSS) approach. 
This method gives the same result that would be obtained without resorting to the mapping, but considerably simplifies the formal approach to the problem. 

We thus proceed by implementing  Schwinger's map that transforms the spin operators into two bosonic operators $a$ and $b$ according to
\begin{equation}
\begin{aligned}
J_z &=& \frac{1}{2}(a^\dagger a - b^\dagger b),\qquad J_+ = (J_-)^\dag= a^\dagger b.	
\end{aligned}
\end{equation} 
To fix the total spin $J$, we impose to work on the restricted subspace where $n_a + n_b = 2J$, with $n_{a,b}$ the expectation value of the number operators for the two Schwinger modes. 
We now introduce standard bosonic coherent states $|\bm{c}\rangle= |\alpha,\beta\rangle$ of such modes and define the corresponding Husimi-Q function as 
\begin{equation}
Q(\alpha,\beta) = \frac{1}{\pi^2} \langle \alpha,\beta | \rho | \alpha,\beta\rangle.
\end{equation}
This Husimi function will also satisfy a quantum Fokker-Planck equation of the form~(\ref{FP}). 
The correspondence table~(\ref{Jp_orbital})--(\ref{Jz_orbital}) now becomes 
\begin{IEEEeqnarray}{rCl}
\label{corr_Jp}[J_+, \rho] &\to& \mathcal{J}_+(Q)  = (\alpha^* \partial_{\beta^*} - \beta \partial_\alpha) Q, 		\\[0.2cm]
\label{corr_Jm}[J_-,\rho] &\to& \mathcal{J}_{-}(Q) = (\beta^* \partial_{\alpha^*} - \alpha \partial_\beta) Q, 		\\[0.2cm]
\label{corr_Jz}[J_z, \rho] &\to& \mathcal{J}_z(Q) = \frac{1}{2}(\alpha^* \partial_{\alpha^*}+  \beta \partial_\beta - \text{c.c}) Q. 		
\IEEEeqnarraynumspace
\end{IEEEeqnarray}

For a single spin-1/2 system, the most general density matrix may be written as 
\begin{equation}\label{rho_spin12}
\rho = \frac{1}{2} (1 + \bm{\tau} \cdot \bm{\sigma}),
\end{equation}
 where $\sigma_i$ are the Pauli matrices and $\tau_i = \tr(\rho \sigma_i)$. 
In this case, it follows that the corresponding Husimi-Q function is given by the particularly simple form \cite{Scully1994}
\begin{equation}\label{Q_spin12}
Q(\alpha,\beta) = \frac{e^{-\bm{c}^\dagger \bm{c}}}{\pi^2} \;  \bm{c}^\dagger \frac{(1 + \bm{\tau} \cdot \bm{\sigma})}{2} \bm{c},
\end{equation}
where $\bm{c} = (\alpha,\beta)$ is to be interpreted as a two-component spinor.
For an arbitrary spin, we write instead
\begin{equation}\label{Q}
Q(\alpha,\beta) = \frac{e^{-\bm{c}^\dagger \bm{c}}}{\pi^2} \; V(\alpha,\beta),
\end{equation}
where
\begin{equation}
V(\alpha,\beta) = \sum\limits_{m,m'}  \frac{\rho_{m,m'} (\alpha^*)^{J+m} (\beta^*)^{J-m} \alpha^{J+m'} \beta^{J - m'}}{\sqrt{(J+m)! (J-m)! (J+m')! (J-m')!}}.
\end{equation}
One may verify that $V(\alpha,\beta)$ is a homogeneous function of degree $2J$ in $\alpha$ and $\beta$. 
Thus, using Euler's theorem for homogeneous functions, we find that
\begin{equation}\label{Euler}
(\alpha \partial_\alpha + \beta \partial_\beta)V(\alpha,\beta) = 2 J V(\alpha,\beta)
\end{equation}
with an identical equation for $\alpha^*$ and $\beta^*$.
\subsection{Relation between the two approaches}
Equation~(\ref{Q}) can be related to the spin coherent state function in Eq.~(\ref{spin_Q}) as follows. 
Define the angle-action variables $\mathcal{I}$, $\theta$, $\phi$, and $\psi$ according to 
\begin{equation}
\label{action_angles}
\begin{aligned}
\alpha &= \sqrt{\mathcal{I}} \cos\frac{\theta}{2} e^{-i (\phi+\psi)/2}, \qquad\beta = \sqrt{\mathcal{I}} \sin\frac{\theta}{2} e^{i (\phi-\psi)/2}.
\end{aligned}
\end{equation}
The integration measure changes as 
\begin{equation}
\ud^2\alpha \ud^2 \beta = \frac{1}{8} \mathcal{I} \ud \mathcal{I}  \ud \psi \ud \Omega.
\end{equation}
After integrating over $\psi$, we obtain 
\begin{equation}\label{eq:vol_element}
\ud^2\alpha \ud^2 \beta = \frac{\pi}{4} \; \mathcal{I} \ud \mathcal{I} \ud\Omega,
\end{equation}
where $\ud \Omega = \sin\theta \ud \theta \ud \phi$.
One may then verify that the Husimi functions~(\ref{Q}) and (\ref{spin_Q}) are related by
\begin{equation}\label{relation_Qs}
Q(\alpha,\beta) = \frac{e^{-\mathcal{I}} \mathcal{I}^{2J}}{\pi^2 (2J)!} \; \mathcal{Q}(\Omega).
\end{equation}
Thus, one may move back and forth between the two representations based on convenience.
Comparing this result with Eq.~(\ref{Q}) also allows us to identify the relation 
\begin{equation}
V(\alpha,\beta) = \frac{\mathcal{I}^{2J}}{(2J)!} \mathcal{Q}(\Omega).
\end{equation}

\subsection{Wehrl entropy}

The entropy associated to the Husimi function is known as the  Wehrl entropy \cite{Wehrl1978,Wehrl1979,Anderson1993,General1995,Zyczkowski2001,DelReal2013a,Gnutzmann2001},
\begin{equation}\label{entropy_spin}
S = - \frac{(2J+1)}{4\pi} \int \ud \Omega \; \mathcal{Q}(\Omega) \ln \mathcal{Q}(\Omega),
\end{equation}
where the constant $(2J+1)/4\pi$ has been introduced only for convenience. 
%The Wehlr entropy measure has been extensively used in the past as an information theoretical tool~\cite{Anderson1993,General1995} and in the characterization of spin phase-space~\cite{Zyczkowski2001,DelReal2013a,Gnutzmann2001,Altland2012}. 

In the TSS representation, the Wehrl entropy may be written as 
\begin{equation}\label{S_main}
S = - \int \ud^2\alpha \ud^2\beta \; Q(\alpha,\beta) \ln Q(\alpha,\beta).
\end{equation}
where both integrals are over the entire complex plane.  
The definitions~(\ref{entropy_spin}) and (\ref{S_main}) are not identical, but differ by an additive constant.
However, in view of Eq.~(\ref{dSdt}),  we will only be interested in the general rate of change of the entropy so we shall not differentiate between the two definitions. 

Unlike the von Neumann entropy, the Wehrl entropy can be affected by unitary transformations. 
This is related to the coarse-graining aspect of the Husimi function.
Hamiltonians which are linear in $J_i$ do not affect $S$, but in general nonlinear Hamiltonians do \cite{Zyczkowski2001}. 
Which classes of Hamiltonians affect the unitary part is still an open question \cite{Wehrl1979}. 
Here we shall not consider this unitary contribution, as it  simply adds a new term to $\ud S/\ud t$,
%. Here we shall not consider any possible unitary contributions
but rather concentrate  on the dissipative contribution to $\ud S/\ud t$, which from Eq.~(\ref{FP}) is found to be
\begin{equation}\label{contribution_D}
\frac{\ud S}{\ud t}\Bigg|_\text{diss} = - \frac{(2J+1)}{4\pi}  \int \ud \Omega \; \mathcal{D}(\mathcal{Q}) \ln \mathcal{Q}.
\end{equation}
The goal  is  to separate this in the form of Eq.~(\ref{dSdt}), i.e., to identify terms that can be interpreted as an entropy production rate $\Pi$ and an entropy flux rate $\Phi$. 

\subsection{Information-theoretic aspects of the Wehrl entropy}

The Wehrl entropy has long been used as an information-theoretic tool when dealing with coherent states. 
Perhaps the most well-developed approach is that of Refs.~\cite{General1995,Buek1995}, where the authors presented an operational interpretation of $S$ in terms of phase-space measurements subject to an additional filtering device (called quantum ruler) that coarse grains the knowledge acquired in the measurement. 
This then leads to the so-called sampling entropies, with the Wehrl entropy representing a special example for the case where the quantum ruler is a coherent state. 
Other operational uses of the Wehrl entropy include nontrivial measures of uncertainty \cite{Anderson1993,Zyczkowski2001}, measures  of localization \cite{Gnutzmann2001}, and its relation to quantum chaos \cite{Altland2012,DelReal2013a}.

\section{\label{sec:dephasing_channel} Dephasing channel}
\subsection{General formulation}
As a first example, we consider the dephasing channel with Lindblad operator 
\begin{equation}\label{D_dephasing}
D(\rho) = - \frac{\lambda}{2} [J_z,[J_z,\rho]].
\end{equation}
This channel does not induce any population changes in the $J_z$ basis, but only causes a loss of coherence.
The corresponding phase-space dissipator is simply 
\begin{equation}
\mathcal{D}(\mathcal{Q}) = - \frac{\lambda}{2}  \mathcal{J}_z (\mathcal{J}_z(\mathcal{Q})), 
\end{equation}
where $\mathcal{J}_z(\mathcal{Q})$ is given in Eq.~(\ref{Jz_orbital}).

By replacing this in Eq.~(\ref{contribution_D}) and integrating by parts, we arrive at
\begin{equation}\label{Pi_dephasing}
\frac{\ud S}{\ud t} = \Pi =\frac{\lambda}{2} \bigg(\frac{2J + 1}{4\pi} \bigg) \int \ud \Omega \; \frac{|\mathcal{J}_z(\mathcal{Q})|^2}{\mathcal{Q}},
\end{equation}
which has the typical form of an entropy production \cite{Tome2010,Spinney2012,Landi2013b,Santos2017b}:
It is always non-negative and zero iff  $\mathcal{J}_z(\mathcal{Q})=0$.
This  occurs only when $\mathcal{Q}$ is independent  of the azimuthal angle $\phi$, which is the phase-space analog of requesting that $\rho$ is diagonal in the $J_z$ basis. 
Hence, this result  establishes $\mathcal{J}_z(\mathcal{Q})$ as the current associated with the loss of coherence in the $J_z$ basis. 

Equation~(\ref{Pi_dephasing}) also shows that a dephasing bath has no associated entropy flux. 
This also appears in the context of the von Neumann entropy production and the Wigner entropy production for bosonic modes \cite{Santos2017b}. 
Moreover, it also agrees with the definition of dephasing as a unital map, for which the entropy can only increase \cite{265894} (whereas $\Pi \geq 0$, the sign of $\Phi$ is in general arbitrary and thus may lead to a reduction in the entropy. But when $\Phi = 0$, we ensure that the entropy can never decrease).  

\subsection{Spin-1/2 case}

We can find an explicit formula for the integral~(\ref{Pi_dephasing}) in the case of  spin-$1/2$ particles [cf. Eqs.~(\ref{rho_spin12}) and (\ref{Q_spin12})]; viz,
\begin{equation}\label{Pi_deph_Wehrl}
\Pi =\frac{\lambda}{4} (\tau_x^2 + \tau_y^2) \bigg\{ \frac{\tau - (1- \tau^2) \tanh^{-1}(\tau)}{\tau^3}\bigg\},
\end{equation}
where $\tau = \sqrt{\tau_x^2 + \tau_y^2 + \tau_z^2}$. 
For a pure state ($\tau\to 1$), we get the particularly simple result
\begin{equation}
\Pi =\frac{\lambda}{4} (\tau_x^2 + \tau_y^2) = \frac{\lambda}{4} \sin^2\theta.
\end{equation}

\begin{figure}
\centering
\includegraphics[scale=1]{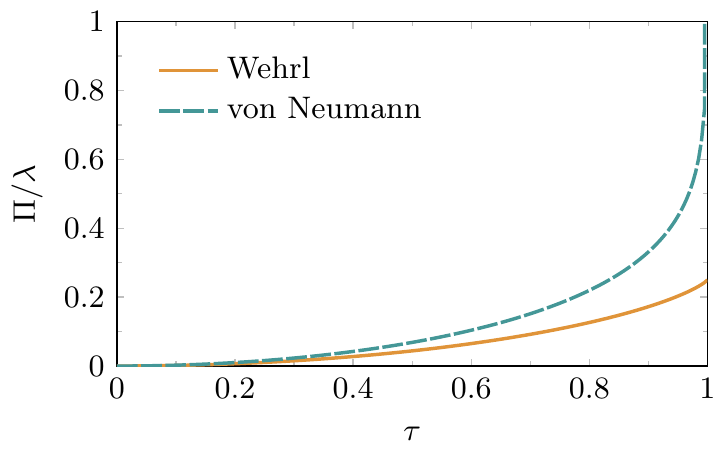}
\caption{\label{fig:dephasing}
The entropy production rate contribution of the dephasing bath for a spin-$1/2$ particle, as a function of $\tau$.
In red we show the von Neumann entropy production rate, given by Eq.~(\ref{Pi_deph_vN}), and in black the corresponding Wehrl entropy production rate, given by Eq.~(\ref{Pi_deph_Wehrl}). 
The curves were computed assuming $\tau_x^2 + \tau_y^2 = \tau^2$. 
The von Neumann measure diverges for a pure state ($\tau \to 1$), whereas the Wehrl measure remains finite.
}
\end{figure}

We can also compare this with the von Neumann formulation in Eq.~(\ref{Pi_vN}).
For the case of a dephasing bath, given by Eq.~(\ref{D_dephasing}), the target state $\rho^*$ will be any diagonal state in the $J_z$ basis.
The entropy production given by Eq.~(\ref{Pi_vN}) is then readily found to be 
\begin{equation}\label{Pi_deph_vN}
\Pi_\text{vN} = \frac{\lambda}{2} \; (\tau_x^2 + \tau_y^2) \frac{\tanh^{-1} (\tau)}{\tau}.
\end{equation}
A comparison of this result for the case where $\tau_x^2 + \tau_y^2 = \tau^2$ is shown in Fig.~\ref{fig:dephasing}. 
As it can be seen, both the Wehrl and the von Neumann entropy productions behave in a similar way. 
However, as the system approaches a pure state ($\tau\to1$), the von Neumann entropy production diverges, whereas the Wehrl entropy production rate remains finite. 
\subsection{Application: spin in a rotating magnetic field}
As an example, let us consider a spin-$1/2$ particle in the presence of a rotating magnetic field.
We take the system Hamiltonian to be 
\begin{equation}\label{H_example}
H(t) = -\frac{b_0}{2}\sigma_z - \frac{b_1}{2}(\sigma_x\cos(\omega t)+\sigma_y\sin(\omega t)),
\end{equation}
and assume that the system is also subject to the dephasing dissipator~(\ref{D_dephasing}). 

\begin{figure}[!htb]
\begin{minipage}[b]{0.40\linewidth}
\includegraphics[width=1\textwidth]{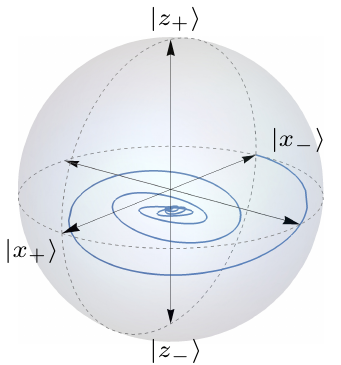}
\end{minipage} \hfill
\begin{minipage}[b]{0.58\linewidth}
\includegraphics[width=1\textwidth]{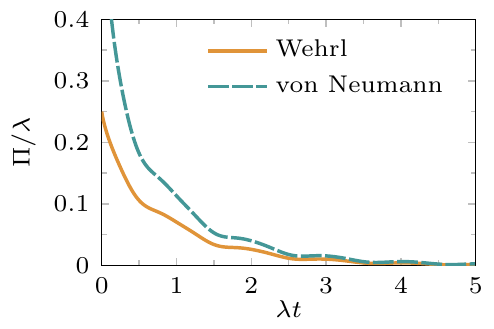}
\end{minipage}
\caption{\label{fig:bloch_dephasing}
Evolution of a spin-$1/2$ particle under a time-dependent magnetic field [Eq.~(\ref{H_example})] and a dephasing bath [Eq.~(\ref{D_dephasing})]. 
Left: trajectory in the Bloch sphere. 
Right: Wehrl and von Neumann entropy production rates. 
We assume the system initially starts in the state $|x_-\rangle = (|z_+\rangle - |z_-\rangle)/\sqrt{2}$ (with $\sigma_z|z_{\pm}\rangle=\pm|z_{\pm}\rangle$).
The chosen parameters were $b_0/\lambda=5$, $b_1/\lambda=1$, and $\omega/\lambda=1$. 
}
\end{figure}

The trajectory of the system in the Bloch sphere is shown in Fig.~\ref{fig:bloch_dephasing}, together with a comparison of the Wehrl and von Neumann entropy production rates. 
As can be appreciated, the Wehrl entropy production rate is capable of capturing the same features as its von Neumann counterpart, but remains finite throughout the motion. 

%
%
%		Amplitude damping
%
%
%%%%%%%%%%%%%%%%%%%%%%%%%%%%%%%%%%%%%%
\section{\label{sec:amplitude_damping} Amplitude damping}

%%%%%%%%%%%%%%%%%%%%%%%%%%%%%%%%%%%%%%
\subsection{Dissipator and relevant currents}

Next we consider the amplitude damping dissipator, which we define as 
\begin{equation}
\begin{aligned}
\label{damp_diss}
D(\rho) &= \gamma(\bar{n}+1) \left[ J_- \rho J_+ - \{ J_+ J_-, \rho\}/2 \right]\\
&+ \gamma \bar{n} \left[ J_+ \rho J_- -  \{ J_- J_+, \rho\}/2\right],
\end{aligned}\end{equation}
where $\bar{n}$ is the mean number of excitations in the environment. 
This dissipator targets the thermal Gibbs state $e^{-\beta H}/\tr[e^{-\beta H}]$  of the Hamiltonian $H = \omega J_z$, provided $\bar{n} = (e^{\beta \omega}-1)^{-1}$. 
When $T  \to 0 $, this state becomes the  ``south-pole''  state $|J,-J\rangle$ when $\omega >0$, and the  ``north-pole'' state $|J,J\rangle$ when $\omega<0$.

It is convenient to define the superoperator
\begin{equation}
f(\rho) = (\bar{n}+1)  \rho J_+ - \bar{n} J_+ \rho
\end{equation}
with which Eq.~(\ref{damp_diss}) can be written as 
\begin{equation}\label{amp_D}
D(\rho) = \frac{\gamma}{2} \left\{ [J_-, f(\rho)] - [J_+, f^\dagger(\rho)]\right\}.
\end{equation}
The super-operator $f(\rho)$ represents a current operator for the density matrix, in the sense that Eq.~(\ref{amp_D}) takes the form of a continuity equation. 
Moreover, one may verify that $f(e^{-\beta H}) = 0$, which allows us to interpret the stationary state as the one for which the current is itself zero. 
Moving to phase space, we have 
\begin{equation}
\mathcal{D}(\mathcal{Q}) = \frac{\gamma}{2}  \bigg\{ \mathcal{J}_{-}(f(\mathcal{Q})) - \mathcal{J}_{+}(f^*(\mathcal{Q})) \bigg\}, 
\end{equation}
where
\begin{equation}\label{fQ_spin}
f(\mathcal{Q}) =  \frac{1}{2} \bigg[2 J \mathcal{Q} - \mathcal{J}_z(Q)\bigg] e^{i \phi} \sin \theta + \frac{1}{2}\bigg[ \cos\theta - (2\bar{n}+1)\bigg] \mathcal{J}_+(\mathcal{Q}),
\end{equation}
[see Eqs.~(\ref{Jp_orbital}) and (\ref{Jm_orbital}) for the definition of the current operators $\mathcal{J}_i$].
Alternatively, in terms of the TSS bosonic representation, the current $f$ becomes
\begin{equation}
\label{fQ}f(Q) = \bigg[ \alpha^* \beta + (\bar{n}+1) \beta \partial_\alpha - \bar{n} \alpha^* \partial_{\beta^*}\bigg] Q	
%\\[0.2cm]
%&=& \frac{e^{-\bm{\alpha}^\dagger \bm{\alpha}}}{\pi^2} \bigg[  (\bar{n}+1) \beta \partial_\alpha - \bar{n} \alpha^* \partial_{\beta^*}\bigg] V
%\label{amp_fV}f(V) &=& \bigg[  (\bar{n}+1) \beta \partial_\alpha - \bar{n} \alpha^* \partial_{\beta^*}\bigg] V
\end{equation}

\subsection{Identification of the entropy production rate}

To separate $\ud S/\ud t$ into the form stated in Eq.~(\ref{dSdt}), we recast all phase-space variables in terms of the relevant current in the problem, which in this case is $f(\mathcal{Q})$.
One then notes that following standard thermodynamic arguments, the entropy production should be an even function of the relevant currents, whereas the entropy flux rate should be odd \cite{Callen1985}. 

It is more convenient to  use Eq.~(\ref{Q}) in order to express $Q$ in terms of the function $V$ since, it turns out,  most differential operators act trivially on the exponential prefactor $e^{-\bm{c}^\dagger \bm{c}}$.
The dissipator then becomes
\begin{equation} \label{contribution_Di}
\mathcal{D}(Q) = \frac{\gamma}{2}\frac{e^{-\bm{c}^\dagger \bm{c}}}{\pi^2}   \left[
\mathcal{J}_{-}(f(V)) - \mathcal{J}_{+}(f^*(V))  \right],
\end{equation}
where
\begin{equation}\label{amp_fV}
f(V) = \left[  (\bar{n}+1) \beta \partial_\alpha - \bar{n} \alpha^* \partial_{\beta^*}\right] V.
\end{equation}

Inserting these currents into Eq.~(\ref{contribution_D}), integrating by parts, and writing everything in terms of $V$ quantities, we get
\begin{equation}\label{tmp3412}
\frac{d S}{dt}\bigg|_\text{diss} =  \frac{\gamma}{2} \int \frac{\ud\bm{c}}{V}\; \frac{e^{-\bm{c}^\dagger \bm{c}}}{\pi^2} \;
%\begin{pmatrix} f(V)^* & f(V) \end{pmatrix}
%\begin{pmatrix} 
%-J_{+}(V) \\ 
%J_{-}(V) 
%\end{pmatrix}
\left[ f(V) \mathcal{J}_-(V) - f(V)^* \mathcal{J}_+(V)\right].
\end{equation}
%Now we  express the currents $\mathcal{J}_\pm$ in terms of $f(V)$, which is accomplished using 
%Eqs.~(\ref{corr_Jp}), (\ref{corr_Jm}) and (\ref{Euler}). We then get
where 
\begin{widetext}
\begin{equation}\label{current_sep}
 f(V) \mathcal{J}_-(V) - f(V)^* \mathcal{J}_+(V) = -  \frac{2 J V (f^* \alpha^* \beta+ f \alpha \beta^*)}{(\bar{n}+1) |\beta|^2 + \bar{n}|\alpha|^2} + \frac{\begin{pmatrix} f^* & f \end{pmatrix} \mathcal{W} \begin{pmatrix} f \\ f^* \end{pmatrix}}{(\bar{n}+1)^2 |\beta|^4 - \bar{n}^2 |\alpha|^4} 
\end{equation}
%where
with
\begin{equation}
\mathcal{W} = \begin{pmatrix}
(\bar{n}+1) |\beta|^4 - \bar{n} |\alpha|^4 	&	(\alpha^* \beta)^2 	\\[0.2cm]
(\alpha\beta^*)^2 	&	(\bar{n}+1) |\beta|^4 - \bar{n} |\alpha|^4
\end{pmatrix}.
\end{equation}
%\end{widetext}
The first term in Eq.~(\ref{current_sep}) is linear in the relevant currents, whereas the second one is quadratic. 
Hence, the first term should naturally be associated with an entropy flux rate and the latter with an entropy production rate. 
That is, we may separate  Eq.~(\ref{tmp3412}) as 
\begin{equation} 
\begin{aligned}
\Pi &= \frac{\gamma}{2} \int \frac{\ud\bm{c}}{V} \frac{e^{-\bm{c}^\dagger \bm{c}}}{\pi^2} \; 
\frac{\begin{pmatrix} f^* & f \end{pmatrix} \mathcal{W} \begin{pmatrix} f \\ f^* \end{pmatrix}}{(\bar{n}+1)^2 |\beta|^4 - \bar{n}^2 |\alpha|^4},\\
%\frac{e^{-\bm{\alpha}^\dagger \bm{\alpha}}}{\pi^2}  
%\bigg\{  
%\frac{2 |f(V)|^2 (|\alpha|^2 + |\beta|^2)}{ (\bar{n}+1) |\beta|^2 + \bar{n} |\alpha|^2} 
% \\[0.2cm]
%&&+|J_z(V)|^2 \bigg[\frac{2 \bar{n} |\alpha|^2}{(\bar{n}+1) |\beta|^2 + \bar{n} |\alpha|^2 }-1\bigg] \bigg\}
%\frac{\gamma}{2} \int \frac{\ud\bm{\alpha}}{V} \; \frac{e^{-\bm{\alpha}^\dagger \bm{\alpha}}}{\pi^2}  
%\begin{pmatrix} 
%f^* & f \end{pmatrix} \mathcal{W} \begin{pmatrix} f \\ f^* \end{pmatrix}
%\\[0.2cm]
\Phi &= \gamma J \int \ud\bm{c} \frac{e^{-\bm{c}^\dagger \bm{c}}}{\pi^2}  \frac{f(V) \alpha \beta^* + f(V)^* \alpha^* \beta}{(\bar{n}+1)|\beta|^2 + \bar{n} |\alpha|^2 }.
\end{aligned}
\end{equation}

We can also express these formulas in terms of angular variables.
%of spin coherent states. 
First, for the entropy flux, we use  Eq.~(\ref{fQ_spin}) and integrate over $\mathcal{I}$ to get
\begin{equation}\label{Phi_damping_angular}
\Phi = \frac{(2J+1)}{4\pi} \gamma J\int\ud \Omega \sin\theta \bigg\{\frac{2 J \mathcal{Q} \sin\theta}{(2\bar{n}+1)-\cos\theta} - \partial_\theta \mathcal{Q}\bigg\}.
\end{equation}
Similarly, for the entropy production, we get 
\begin{equation} \label{eq:wehrl_prod_ad_def}
\Pi = \frac{\gamma(2J+1)}{8\pi}\int\frac{\ud \Omega}{\mathcal{Q}} \frac{\begin{pmatrix} f(\mathcal{Q})^* & f(\mathcal{Q}) \end{pmatrix} \tilde{\mathcal{W}} \begin{pmatrix} f(\mathcal{Q}) \\ f(\mathcal{Q})^* \end{pmatrix}	}{(\bar{n}+1)^2 \sin^4 (\theta/2) - \bar{n}^2 \cos^4(\theta/2)},
%\\[0.2cm]
%\Phi_a &=& \frac{\gamma J}{4\pi} \int\ud \Omega \sin\theta \bigg\{\frac{2 J \mathcal{Q} \sin\theta}{(2\bar{n}+1)-\cos\theta} - \partial_\theta \mathcal{Q}\bigg\}
\end{equation}
where $\tilde{\mathcal{W}} = \mathcal{W}/\mathcal{I}^2$. 
%%This expression for the entropy production rate is still somewhat cumbersome and does not allow for a clear identification of the relevant currents. 
%% We can obtain a physically more significant formula by noting that, since 
As $V$  is a homogeneous function [Eq.~(\ref{Euler})], one may relate the currents $f(V)$ and $\mathcal{J}_z(V)$ according to
\begin{equation}\label{Jz_f}
\mathcal{J}_z(V) = \frac{f(V)^* \alpha^* \beta-f(V) \alpha \beta^*}{(\bar{n}+1) |\beta|^2 - \bar{n} |\alpha|^2}.
\end{equation}
Using this result and expanding the matrix $\tilde{\mathcal{W}}$, we then arrive at
%\begin{widetext}
\begin{equation}\label{Pid}
%\Pi_d = \frac{\gamma}{2} \frac{(2J+1)}{4\pi} \int \frac{\ud \Omega}{\mathcal{Q}} \frac{[f(\mathcal{Q}) e^{-i \phi} + f(\mathcal{Q})^* e^{i \phi}]^2}{(2\bar{n}+1)-\cos\theta}
\Pi = \frac{\gamma}{2} \frac{(2J+1)}{4\pi} \int \frac{\ud \Omega}{\mathcal{Q}} 
\Bigg\{ 
\frac{[
2 J \mathcal{Q} \sin\theta + (\cos\theta - (2\bar{n}+1)) \partial_\theta \mathcal{Q}
]^2}{(2\bar{n}+1)-\cos\theta}
+ 
 |\mathcal{J}_z(\mathcal{Q})|^2 \bigg[(2\bar{n}+1)\cos\theta-1\bigg] \frac{\cos\theta}{\sin^2\theta}
 \Bigg\}.
\end{equation}
\end{widetext}
The most striking feature about this result is the appearance of the dephasing current $|\mathcal{J}_z(\mathcal{Q})|^2$  [Eq.~(\ref{Pi_dephasing})] as a part of the entropy production. 
This means that the Wehrl entropy production rate is able to capture the contribution of the  amplitude damping to the loss of coherence. 
Thus, not only do we get a microscopic picture of the irreversible currents responsible for the entropy production, but we are also able to distinguish the different contributions related to the amplitude damping current $f$ and the dephasing current $\mathcal{J}_z$.
Moreover, compared to Eq.~(\ref{Pi_dephasing}), we also see a temperature-dependent prefactor multiplying $|\mathcal{J}_z(\mathcal{Q})|^2$.
This term introduces an angular dependence  of the dephasing and is a consequence of the fact that for the amplitude damping, decoherence is not homogeneous over the Bloch sphere.

\subsection{Spin-1/2 case}

We now consider the case of spin $1/2$, where all formulas become much simpler. 
We express all results in terms of the bath-induced magnetization  $\bar{\tau}_z = -1/(2\bar{n}+1)$. 
The energy flux~(\ref{PhiE_damp}) simplifies to 
\begin{equation}
\Phi_E = \frac{\gamma \omega}{2\bar{\tau}_z} (\bar{\tau}_z - \tau_z ).
\end{equation}
The entropy flux~(\ref{Phi_damping_angular}), on the other hand, simplifies to 
\begin{equation}\label{Phi_damp_W}
\Phi  = \frac{\nicefrac{\gamma}{2}}{\bar{\tau}_z^3}
\bigg(
\bar{\tau}_z+(\bar{\tau}_z^2-1)\tanh^{-1}(\bar{\tau}_z)
\bigg)(\tau_z-\bar{\tau}_z),
\end{equation}
They are related by
\begin{equation}
\Phi = \bigg[\frac{(1-\bar{\tau}_z^2)\tanh^{-1}(\bar{\tau}_z)-\bar{\tau}_z}{\bar{\tau}_z^2\; \omega}\bigg]\Phi_E.
\end{equation}
In the limit  $T\gg\omega$, this becomes approximately
\begin{equation}
\Phi\simeq\frac{1}{3}\frac{\Phi_E}{T},
\end{equation}
which is a particular case of Eq.~(\ref{Phi_Phi_E_high_T}).

Finally, we present the result for the entropy production rate, given by Eq.~(\ref{Pid}):
\begin{equation}\label{Pi_damp_spin12}
\Pi = \Phi + \frac{\gamma}{2}
\frac{2\bar{\tau}_z\tau_z-(\tau^2+\tau_z^2)}{2\bar{\tau}_z}
\;\bigg\{\frac{\tau-(1-\tau^2)\tanh^{-1}(\tau)}{\tau^3}\bigg\}
\end{equation}
where $\Phi$ is given by Eq.~(\ref{Phi_damp_W}).

\begin{figure}[h!]
\centering
\includegraphics[width=0.4\textwidth]{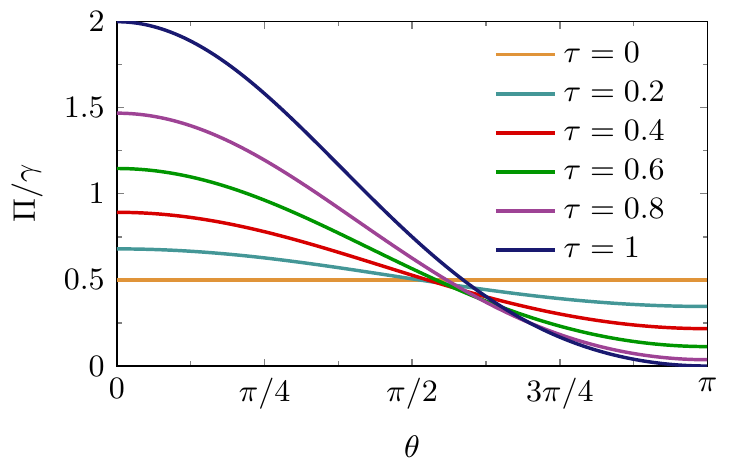}
\caption{\label{fig:damping_theta}
The Wehrl entropy production rate~(\ref{eq:Prod.damp.T0}) at $T = 0$ as a function of $\theta$, where we parametrize $\bm\tau=(\tau \sin \theta,0,\tau \cos\theta)$. The curves correspond to different values of $\tau$ and therefore illustrate the behavior as one goes from a maximally mixed state ($\tau = 0$) to a pure state ($\tau = 1$).
}
\end{figure}

For comparison, the von Neumann entropy flux rate [Eq. (\ref{Phi_vN})] is 
\begin{equation}\label{Phi_damp_vN}
\Phi_{\text{vN}}  =  \gamma\frac{\tanh^{-1}(\bar{\tau}_z)}{\bar{\tau}_z}(\tau_z-\bar{\tau}_z)
\end{equation}
whereas the entropy production rate~(\ref{Pi_vN}) reads
\begin{equation}\label{Pi_damp_vN}
\Pi_{\text{vN}}  =  \Phi_{\text{vN}} - 
\frac{\gamma}{2}\frac{\tanh^{-1}(\tau)}{\tau\bar{\tau}_z}
\bigg( \tau^2 + \tau_z(\tau_z - 2\bar{\tau}_z)\bigg)
\end{equation}
Note how $\Phi_\text{vN}$ diverges in the limit $\bar{\tau}_z \to - 1$, in agreement with Eq.~(\ref{Phi_vN}).
\begin{figure}[t!]
\centering
\includegraphics[width=0.4\textwidth]{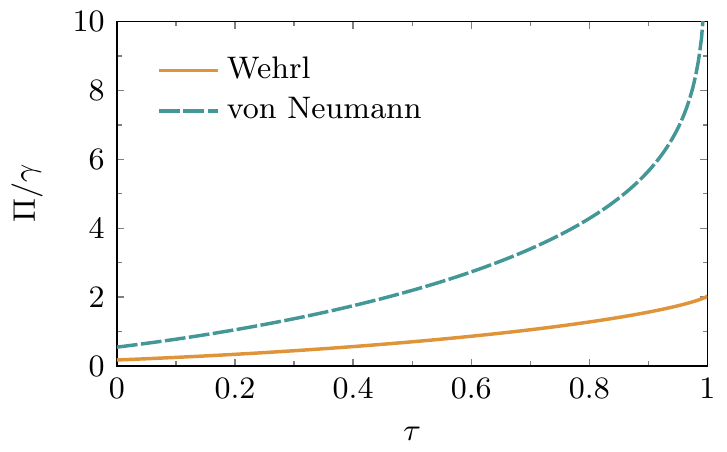}
\caption{\label{fig:damping}
The entropy production rate [von Neumann given by Eq.~(\ref{Pi_damp_vN}) and Wehrl given by Eq.~(\ref{Pi_damp_spin12})] contribution of the amplitude bath for a spin-$1/2$ particle, as a function of $\tau = \tau_z$ for $\bar{\tau}_z = -1/2$.
}
\end{figure}
In fact, these measures diverge both when the bath is at zero temperature and when the state of the system is pure.
A comparison between the von Neumann and Wehrl entropy production rates for the amplitude damping is shown in Fig.~\ref{fig:damping}.
In the limit $T \to 0$ ($\bar{\tau}_z \to -1$), the Wehrl entropy production and fluxes become
\begin{IEEEeqnarray}{rCl}
\label{eq:Flux.damp.T0}
\Phi &=&  \frac{\gamma}{2}(1+\tau_z)	\\[0.2cm]
\label{eq:Prod.damp.T0}
\Pi &=& 
\Phi+
\gamma\frac{\tau^2 + \tau_z(2+\tau_z)}{4\tau^3}[\tau+(\tau^2-1)\tanh^{-1}(\tau)] \quad\quad
\end{IEEEeqnarray}
The behavior of the entropy production $\Pi$ in the limit $T\to 0 $ is shown in Fig.~\ref{fig:damping_theta}. 

\subsection{Wehrl entropy flux for a general spin}

In the Appendix, we compute the entropy flux given by Eq.~(\ref{Phi_damping_angular}) exactly for a general spin. We also show that the Wehrl entropy flux for $T\rightarrow 0$ simplifies to 
\begin{equation}
\Phi =  2  \gamma  J [  J +  \langle J_z \rangle ],
\end{equation}
which is valid for any $J$. 

Thus, as with the dephasing noise, with the Wehrl formalism, we obtain a well-behaved result even at zero temperature. 
The structure of this expression is also surprisingly similar to the structure found for bosonic systems in Ref.~\cite{Santos2017b}.
It shows that the flux is related to the difference between the instantaneous  value of $\langle J_z \rangle$ and the bath-induced value $-J$ (which is the target state of the amplitude damping at $T\to0$).

\subsection{Energy flux vs Entropy flux}

We can also relate the entropy flux with the energy flux, assuming a Hamiltonian $H=\omega J_z$.
The energy flux is given by $\Phi_E=-\tr(HD(\rho))$, which may be written as 
\begin{IEEEeqnarray}{rCl}
%\Phi_E &=& \frac{\gamma\omega}{\bar{\tau}_z}\sum_{m=-J}^{J} \rho_{m,m}\bigg[  \bar{\tau}_zJ(J+1)  - m(1+m\bar{\tau}_z) \bigg]
%\\[0.2cm]
\label{PhiE_damp}\Phi_E &=& \frac{\gamma\omega}{\bar{\tau}_z}\bigg[  \bar{\tau}_z[J(J+1)-\langle J_z^2 \rangle]  - \langle J_z \rangle  \bigg].
\end{IEEEeqnarray}
In general, even though both $\Phi$ and $\Phi_E$ only depend on the diagonal entries of $\rho$, they are not directly proportional to each other. 
However, taking the limit where $T\gg\omega$, we may approximate Eq.~(\ref{eq:phi_damp_Q3}) to
\begin{equation}\label{Phi_Phi_E_high_T}
\Phi \simeq\frac{1}{(1+1/J)}\frac{\Phi_E}{T}.
\end{equation}
If we then take both  $T\gg\omega$ and $J\rightarrow\infty$, we recover the von Neumann result~(\ref{Phi_vN}).
Thus, in the classical limit, we recover the usual thermodynamic results of the von Neumann framework, which is a key consistency requirement of such a theory. 

%%%%%%%%%%%%%%%%%%%%%%%%%%%%%%%%%%%%%%
%
%
%		APPLICATIONS	
%
%
%%%%%%%%%%%%%%%%%%%%%%%%%%%%%%%%%%%%%%
\section{\label{sec:applications} Applications}

%%%%%%%%%%%%%%%%%%%%%%%%%%%%%%%%%%%%%%
\subsection{Spontaneous emission}

\begin{figure*}[!t]
\centering
\includegraphics[width=0.3\textwidth]{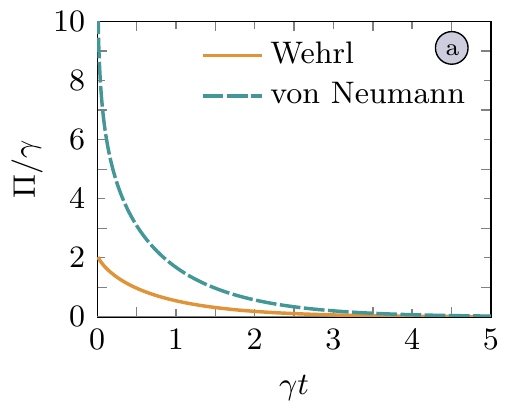}\hfill
\includegraphics[width=0.3\textwidth]{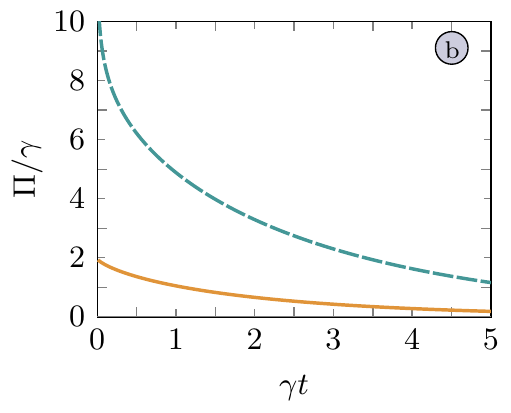}\hfill
\includegraphics[width=0.3\textwidth]{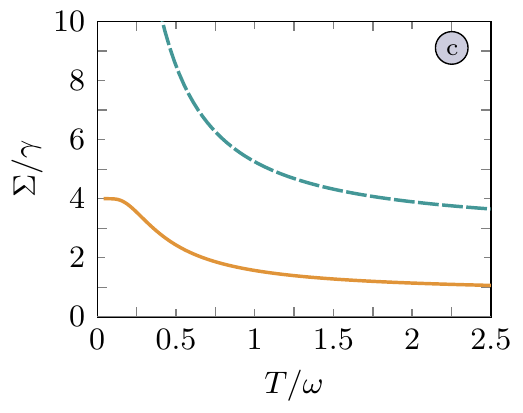}\hfill
\caption{\label{fig:spont_emiss}
Entropy produced during spontaneous emission of an atom starting in the excited state and subject to the amplitude damping dissipator. 
(a) $T/\omega = 1.0$ and (b) $T/\omega = 0.2$.
(c) The total entropy production~(\ref{Sigma}) as a function of temperature.
}
\end{figure*}

We now present several applications of the amplitude damping results, focusing on the case of spin-1/2 particle.
We start with the case of spontaneous emission.
In Figs.~\ref{fig:spont_emiss}(a) and \ref{fig:spont_emiss}(b), we show the entropy production as a function of time for a system starting in the excited state, for two different temperatures. 
As can be seen, as the temperature goes down, the von Neumann entropy production gradually starts to diverge, whereas the Wehrl entropy production rate remains well behaved. 
We also study the total entropy produced in the process, defined as
\begin{equation}\label{Sigma}
\Sigma = \int\limits_0^\infty  \Pi(t)\ud t.
\end{equation}
Results as a function of temperature are shown in Fig.~\ref{fig:spont_emiss}(c). 

\subsection{A thermal quench}

Now we consider a thermal quench. We assume the bath temperature is $T$, but the system begins in a different temperature $T_0$. 
The exact solution of the Lindblad master equation will continue to be a thermal Gibbs state, but with a time-dependent temperature $\beta(t)$.
It is more convenient to work with $\tau_z(t)=-\tanh(\omega\beta(t)/2)$.
Since $\tau_z(t)=\langle \sigma_z \rangle_t$, it follows that this quantity satisfies the differential equation
\begin{equation}
\frac{d\tau_z(t)}{dt} = \frac{\gamma}{\bar{\tau}_z}(\tau_z(t)-\bar{\tau}_z),
\end{equation}
whose solution is
\begin{equation}\label{quench_sol}
\tau_z(t) = \bar{\tau}_z + e^{-\gamma t/|\bar{\tau}_z|}(\tau_z(0)-\bar{\tau}_z).
\end{equation}

\begin{figure}[h!]
\centering
\includegraphics[scale=1]{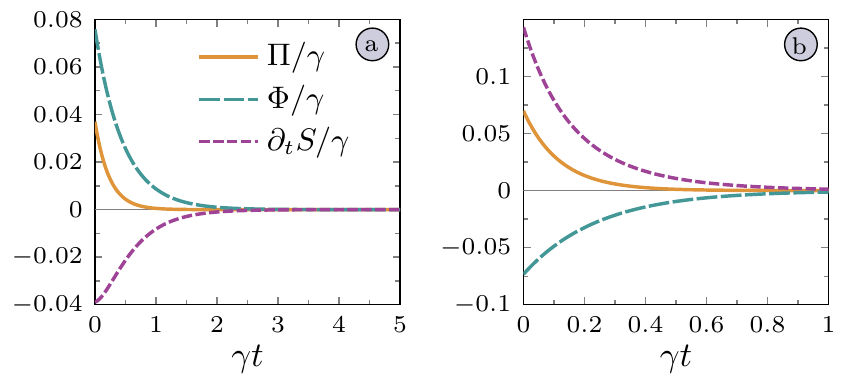}
\caption{\label{fig:thermal}
The Wehrl entropy production rate [Eq.~(\ref{Pi_damp_spin12})], the entropy flux rate [Eq.~(\ref{Phi_damp_W})] and the total rate of change of the entropy, [Eq.~(\ref{dSdt})], for a thermal quench, computed using Eq.~(\ref{quench_sol}) for (a) $T/\omega = 1$, $T_0/\omega = 2$ and (b) the inverse.
}
\end{figure}

Thus, we may readily apply Eqs.~(\ref{Phi_damp_W}) and (\ref{Pi_damp_spin12}) to compute the entropy productions and fluxes for both the von Neumann and the Wehrl entropies. 
Examples of these curves are shown in Fig.~(\ref{fig:thermal}).

\subsection{Spin-$1/2$ in an oscillating magnetic field}

\begin{figure}[!htb]
\begin{minipage}[t]{0.40\linewidth}
\includegraphics[width=1\textwidth]{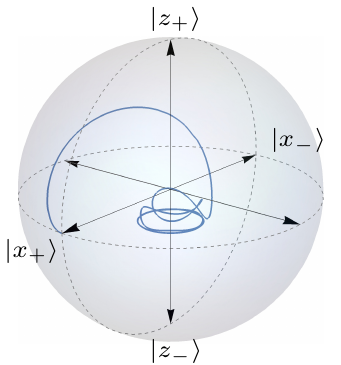}
\end{minipage} \hfill
\begin{minipage}[b]{0.58\linewidth}
\includegraphics[width=1\textwidth]{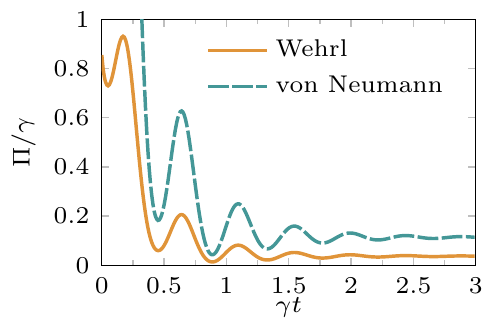}
\end{minipage}
\caption{\label{fig:bloch_damping}
Evolution of a system under a time-dependent magnetic field [Eq.~(\ref{H_example})] and subject to the amplitude damping dissipator [Eq.~(\ref{damp_diss})].
Left: evolution of the spin in the Bloch sphere, starting in the $x$ direction.
Right:  Wehrl and von Neumann entropy production rates. 
In this example, we are considering $b_0/\gamma=5$, $b_1/\gamma=10$, $\omega/\gamma=5$, and $\bar{\tau}_z=-1/3$.
}
\end{figure}

Here we consider again the problem defined by the Hamiltonian [Eq.~(\ref{H_example})], but now subject to the amplitude damping dissipator.
We assume that the system starts initially in the eigenstate of $\sigma_x$ with eigenvalue $+1$. 
To illustrate the physics of the problem, in Fig.~\ref{fig:bloch_damping} we show the dynamics of the spin using the Bloch sphere. 

In the long-time limit, the system tends to its stationary state where $\ud S/\ud t= 0$ so $\Pi= \Phi$. 
For the von Neumann case, we get, in the steady state,
\begin{equation}
\Pi_\text{vN} = \Phi_{\text{vN}}=   -\frac{ 2\gamma b_1^2\bar{\tau}_z^2  \tanh^{-1}(\bar{\tau}_z)}{\gamma^2+2\bar{\tau}_z^2(b_1^2+2(b_0+\omega)^2)}.
\end{equation}
For the Wehr entropy, we have instead
\begin{equation}
\Pi = \Phi= - \frac{\gamma b_1^2 [\bar{\tau}_z+(\bar{\tau}_z^2-1)\tanh^{-1}(\bar{\tau}_z)]  }{\gamma^2+2\bar{\tau}_z^2(b_1^2+2(b_0+\omega)^2)}.
\end{equation}
In the limit  $T\rightarrow 0$ the former diverges, whereas the latter tends to 
\begin{equation}
\Pi =\frac{\gamma b_1^2}{\gamma^2 + 2b_1^2 + 4(b_0+\omega)^2}.
\end{equation}
%The corresponding energy flux is given by Eq.~(\ref{Phi_basic}):
%\begin{IEEEeqnarray}{rCl}
%\Phi_{E} &=&  \frac{\gamma \omega  b_1^2\bar{\tau}_z^2}{\gamma^2+2\bar{\tau}_z^2(b_1^2+2(b_0+\omega)^2)} = T \Phi_{a,{\text{vN}}} 
%%\nonumber \\
%%&=& T \Phi_{a,{\text{vN}}}
%\end{IEEEeqnarray}
%
%with $T=-\omega/(2\tanh^{-1}(\bar{\tau}_z))$.
%
%%%%%%%%%%%%%%%%%%%%%%%%%%%%%%%%%%%%%%
%
%
%		CONCLUSIONS	
%
%
%%%%%%%%%%%%%%%%%%%%%%%%%%%%%%%%%%%%%%

%
\subsection{\label{sec:fred} Excitation of a two-level atom by a single photon}
Finally, we consider the interaction of a two-level atom with a quantized propagating pulse in free space. The Hamiltonian in this case can be written as (in the interaction picture)
\begin{align}
H= -i\sum_n \big[ 
g_n  \sigma_+a_n         e^{+i\Delta_nt}-
g_n^*\sigma_-a_n^\dagger e^{-i\Delta_nt} \big],
\end{align}
with $\Delta_n=(\omega_0-\omega_n)$.
Here, $\omega_0$ is the atom frequency, $\omega_n$ is the mode frequency and $g_n$ is the coupling constant.
The dynamics restricted to the one excitation in the system is described by the state vector 
\begin{align}
|\psi(t)\rangle = a(t) |e,0\rangle+\sum_n b_n(t)|g,1_n\rangle.
\end{align}
Here $|e\rangle$ ($|g\rangle$) is the excited (ground) state of the atom.
Let us consider the pulse mode to be a single-photon wave packet, which can be written as \cite{PhysRevA.83.063842,PhysRevA.86.053825}
\begin{align}
|1_p\rangle = \sum_n g_n^* f(\omega_n) |1_n\rangle.
\end{align}
From the Schr\"{o}edinger equation $\partial_t|\psi(t)\rangle=-iH|\psi(t)\rangle$ we obtain,
\begin{align}
\partial_t a(t) &= -\sum_n g_n b_n(t) e^{-i\Delta_nt}, \\
\partial_t b_n(t) &= g_n^* a(t)e^{i\Delta_nt},
\end{align}
By formally integrating $b_n(t)$, substituting it in the equation for $a(t)$ and doing a Wigner-Weisskopf approximation in the continuum limit \cite{PhysRevA.83.063842,PhysRevA.86.053825}, we obtain
\begin{align} \label{eq:a}
a(t) &= a(t_0)e^{-\gamma_0(t-t_0)/2} \nonumber\\ &-\sqrt{\gamma_0}e^{-\gamma_0t/2}\int_{t_0}^t\ud t' \xi(t') e^{(\nicefrac{\gamma_0}{2}+i(\omega_0-\omega_p))t'},
\end{align}
where $\xi(t)$ is the wave function for the pulse bandwidth,
\begin{align}
\xi(t) = \frac{\sqrt{\gamma_0}}{2\pi}\int\ud\omega_k f(\omega_k) e^{-i(\omega_k-\omega_p)t} ,
\end{align}
with $\omega_p$ being the central pulse frequency and $\gamma_0$ the standard spontaneous decay rate in free space.

The non-unitary time evolution of the atom can also be approached by using a master equation. 
Considering the atomic density operator $\rho(t)=\tr_{\text{field}}\{|\psi(t)\rangle\langle\psi(t)|\}$, we can write the master equation for the atom in the form \cite{Drobny,Brito}
\begin{align} \label{eq:masterequation}
\partial_t \rho = -i[H,\rho] + \Gamma_t[\sigma_{-}\rho\sigma_+-\{\sigma_+\sigma_-,\rho\}/2],
\end{align}
where $H=\omega_t\sigma_z/2$ with  
\begin{align}
\omega_t &= -\im\bigg[\frac{\partial_ta(t)}{a(t)}\bigg], \\
\Gamma_t &= -2\re\bigg[\frac{\partial_ta(t)}{a(t)}\bigg].
\end{align}

By using Eq.~(\ref{eq:a}), we can write  $\Gamma_t$ as
\begin{align}
\Gamma_t = \gamma_0+\frac{1}{2} \frac{\sqrt{\gamma_0}}{|a(t)|^2}\re\{a^*(t)\xi(t)e^{i(\omega_0-\omega_p)t}\}. 
\end{align}

\begin{figure}[t!]
\centering
\includegraphics[scale=1]{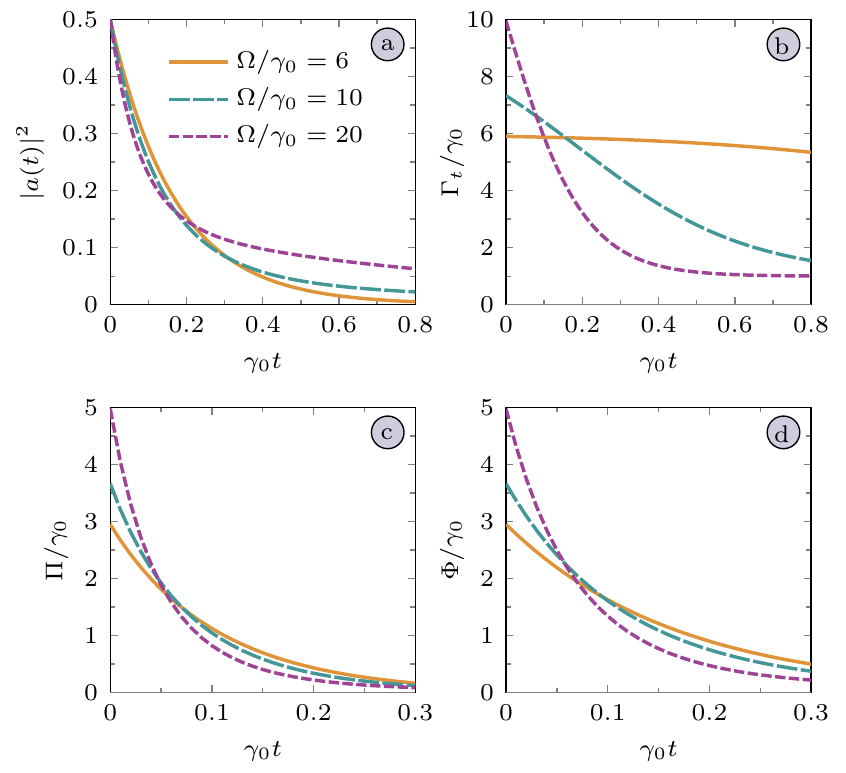}
\caption{\label{fig:pulse}
Two-level atom interacting with a single-photon pulse for different values for $\Omega/\gamma_0$. Here we are considering $a(0)=\sqrt{0.5}$. (a) Excitation probability $|a(t)|^2=\tr\{\rho(t) |e\rangle \langle e|\}$. (b) Effective decay constant $\Gamma_t$. (c) Wehrl entropy production. (d) Wehrl entropy flux.
}
\end{figure}

Note now that we can apply the formalism established in Sec. \ref{sec:amplitude_damping} in order to quantify the Wehrl entropy production for the dynamics given by Eq.~(\ref{eq:masterequation}). 
As an example, let us consider a exponentially decaying pulse 
\begin{align}
\xi(t)=\left \{
\begin{array}{ccc}
N\sqrt{\Omega}\exp(-\Omega t/2), & \text{for} & t>0\\
0, & \text{for} & t<0
\end{array} \right.
\end{align}
with normalization $N=\sqrt{1-|a(t_0)|^2}$. Here $\Omega$ is the pulse bandwidth \cite{PhysRevA.83.063842}.
If we consider $\omega_p=\omega_0$ and $\Omega>\gamma_0$, it is possible to show that the condition 
\begin{align}
a(0)\geq\sqrt{\frac{\delta}{1+\delta}}, \ \ \text{with} \ \ \delta=\frac{4\Omega/\gamma_0}{(1-\Omega/\gamma_0)^2} ,
\end{align}
ensures $\Gamma_t \geq0$. This way the dynamics will always be Markovian \cite{PhysRevA.89.042120}. 
Thus, we may readily apply Eqs.~(\ref{Phi_damp_W}) and (\ref{Pi_damp_spin12}) to compute the entropy productions and fluxes for the Wehrl entropy [note here that $\bar{\tau}_z=-1$, $\tau=\tau_z=2|a(t)|^2-1$ and $\gamma\rightarrow\Gamma_t$]. 
Examples of these curves are shown in Fig.~\ref{fig:pulse}.
\section{\label{sec:conclusions} Conclusions}

We have put forward a theory for the entropy production of a open quantum spin system based on the Wehrl entropy. 
To date, there is no self-consistent theory of entropy production formulated for arbitrarily dimensional Hilbert spaces. With the proposed theory we take a step to fill this gap.
The applications that we have discussed, including dephasing and amplitude damping baths, show both the potential of the proposed approach and the breath of physically relevant situation that it is able to address. 
We have made the connection between the Wehrl entropy flux rate, Eq.~(\ref{eq:phi_damp_Q3}), and the Clausius relation between entropy and heat, Eq.~(\ref{Phi_vN}), and verified that the former tends to the latter in the classical limit [see Eq.~(\ref{Phi_Phi_E_high_T})].
Looking ahead, we hope that the methods presented here can be used and extended to study entropy production in other physical models involving, for instance, quantum chaos and equilibration for unitary quantum dynamics \cite{PhysRevLett.108.073601,1367-2630-14-7-073011}.

\begin{acknowledgements}  

FB is supported by the Instituto Nacional de Ci\^encia e Tecnologia - Informa\c{c}\~ao Qu\^antica (INCT-IQ).
GTL would like to acknowledge the S\~ao Paulo Research Foundation, under grant number 2016/08721-7. JPS would like to acknowledge the financial support from the CAPES (PNPD program) for the postdoctoral grant. LCC acknowledges support from CNPq (Grants No. 401230/2014-7, 305086/2013-8 and 445516/2014-3). MP acknowledges support from the EU Collaborative project TherMiQ (grant agreement 618074), the DfE-SFI Investigator Programme (grant 15/IA/2864) and the Royal Society Newton Mobility Grant NI160057. 

\end{acknowledgements}  

\appendix*

\section{\label{Exactformula} Exact formula for the entropy flux}

The integrals in Eqs.~(\ref{Phi_damping_angular}) and (\ref{Pid}) must be computed numerically. 
Fortunately, though, Eq.~(\ref{Phi_damping_angular}) for the entropy flux can be computed exactly for general spin and expressed in terms of an arbitrary density matrix $\rho=\sum_{m,m'}\rho_{m,m'}|m\rangle\langle m'|$.
The result reads
\begin{widetext} 
\begin{IEEEeqnarray}{rCl}\label{eq:phi_damp_Q3}
\Phi &=& 
\gamma J \bigg\{ \frac{1+\bar{\tau}_z}{\bar{\tau}_z}+2(J+\langle J_z \rangle)- 
\frac{1}{2}\bigg(\frac{1+\bar{\tau}_z}{1+J}\bigg)
\sum_{m=-J}^J 
\rho_{m,m}\bigg[
%\nonumber
%\\[0.2cm] & &
\bigg(\frac{1+J-m}{ \bar{\tau}_z }\bigg)\; {_2F}\bigg(1,1+J+m\; ; 3+2J \; ; \frac{2\bar{\tau}_z}{\bar{\tau}_z-1}\bigg)
\nonumber
\\[0.2cm]
&+&
\frac{(1+J+m)(1+4J+1/\bar{\tau}_z)}{1-\bar{\tau}_z} \; {_2F}\bigg(1,2+J+m\; ; 3+2J \; ;\frac{2\bar{\tau}_z}{\bar{\tau}_z-1}\bigg)
\bigg] \bigg\}, 
\end{IEEEeqnarray}
where ${_2F}(\cdot)$ is the Gauss hypergeometric function and we have defined
\begin{equation}
\bar{\tau}_z = -\frac{1}{(2\bar{n}+1)}.
\end{equation}  
This is the bath-induced magnetization for a spin 1/2 system (although Eq.~(\ref{eq:phi_damp_Q3}) holds for arbitrary spin).
Note how the entropy flux depends only on the diagonal elements of the density matrix.

When $T\to 0 $ ($\bar{\tau}_z\rightarrow -1$), this result simplifies dramatically to 
\begin{equation}
\Phi =  2  \gamma  J [  J +  \langle J_z \rangle ],
\end{equation}
which is valid for any $J$. 

Eq.~(\ref{eq:phi_damp_Q3}) can also be simplified for the case where $J$ is large and/or $\bar{\tau}_z$ is small.
In this case, using the asymptotic expansion of Ref.~\cite{Ferreira2006}, we get
\begin{equation} \label{eq:simplified}
%\begin{aligned}
\Phi \approx
2 \gamma J \bigg\{
J+\langle J_z \rangle 
+
\frac{1+\bar{\tau}_z}{2\bar{\tau}_z}
\bigg[1 - \frac{3+2J}{2(1+J)}\bigg\langle
 \frac{(1+J+\langle J_z \rangle)(1+(1+4J)\bar{\tau}_z)}{3+2J+(2\langle J_z \rangle+1)\bar{\tau}_z}
-\frac{(1+J-\langle J_z \rangle)(\bar{\tau}_z-1)}{3+2J+(2\langle J_z \rangle-1)\bar{\tau}_z}
\bigg\rangle
\bigg]\bigg\},
%\end{aligned}
\end{equation}
\end{widetext}
which becomes exact in the limit $J\rightarrow\infty$ and/or $\bar{\tau}_z\rightarrow 0$.
Notwithstanding, we find that it gives a remarkably good approximation also for moderately small $J$.

%\bibliography{/Users/gtlandi/Documents/library}
\bibliography{library}
\end{document}